\documentclass[twocolumn,showpacs,showkeys,preprintnumbers,superscriptaddress,amsmath,amssymb]{revtex4}
\usepackage[dvips]{graphicx}
\usepackage{latexsym,amssymb,amsmath,epsfig,bm,times,psfrag}
\usepackage{color}
\usepackage[bookmarksnumbered,bookmarksopen,colorlinks,citecolor=red,linkcolor=blue]{hyperref}
\usepackage{epstopdf}
\newcommand{\Fig}[1]{Fig.~\ref{#1}}
\newcommand{\eq}[1]{Eq.~(\ref{#1})}

\renewcommand{\part}{{\rm part}}

\newcommand{\be}{\begin{equation}}
\newcommand{\ee}{\end{equation}}
\newcommand{\bear}{\begin{eqnarray}}
\newcommand{\eear}{\end{eqnarray}}
\newcommand{\ba}{\begin{array}}
\newcommand{\ea}{\end{array}}

\begin{document}

\title{Impact of spin polarization on transport and thermodynamic coefficients}

\author{De-Xian Wei}
\email{dexianwei@gxust.edu.cn}
\affiliation{School of Science, Guangxi University of Science and Technology, Liuzhou, 545006, China}
\date{\today}

\begin{abstract}
This work investigates the influence of parton spin polarization on effective transport and thermodynamic coefficients in noncentral light- and heavy-ion collisions. To model this influence, I consider two sources of spin polarization: thermal vorticity, induced by angular momentum, and thermal shear, arising from local velocity gradients.
Using a novel kinetic theory framework, one finds that transport and thermodynamic coefficients--including the speed of sound squared $c_{s}^{2}$, specific shear viscosity $\eta/s$, specific bulk viscosity $\zeta/s$, and mean free path $\lambda$--are substantially modified by spin polarization effects.
Among the two sources, thermal vorticity-induced spin polarization dominates the modifications to these coefficients. Moreover, both $c_{s}^{2}$ and $\zeta/s$ exhibit a nonmonotonic dependence on the collision energy, and the associated scaling behaviors potentially serve as indicators of the critical phenomena of quantum chromodynamics.
\end{abstract}



\maketitle

\section{Introduction}
\label{introduction}

Heavy-ion collision experiments at the BNL Relativistic Heavy Ion Collider (RHIC) and the CERN Large Hadron Collider (LHC) have created, in
Au+Au~\cite{BRAHMS:2004adc,PHOBOS:2004zne,STAR:2005gfr,PHENIX:2004vcz} and Pb+Pb~\cite{Muller:2012zq} collisions, a hot and dense medium of strongly interacting matter known as the quark-gluon plasma (QGP). Recently, key QGP-like signatures, such as collective flow-previously observed in large systems have also been detected in small systems, including p+Pb collisions at the LHC~\cite{CMS:2015yux}, and p+Au, d+Au, and $^3$He+Au collisions at the RHIC~\cite{PHENIX:2018lia,Nagle:2018nvi}. However, there remains no conclusive evidence for QGP formation in small systems from hard probe modifications, such as high-$p_T$ jet quenching~\cite{Nagle:2018nvi,CMS:2016xef,Grosse-Oetringhaus:2024bwr}.
If the formation of a QGP droplet in small systems is confirmed, a systematic investigation across system sizes becomes essential. This raises a fundamental question: What is the smallest size for a droplet of strongly coupled hydrodynamic matter? Addressing this requires a comprehensive analysis using effective frameworks that incorporate both realistic initial conditions and medium transport properties to model the space-time evolution of the system.
In recent years, phenomenological studies in relativistic heavy-ion collisions have placed constraints on the transport properties of the QGP, particularly on its specific viscosities. First-principles calculations of these properties remain challenging due to the fermion sign problem and the strongly coupled nature of the QGP~\cite{Troyer:2004ge,Altenkort:2022yhb}. Nonetheless, a range of effective models, including Yang-Mills theory~\cite{Haas:2013hpa}, Polyakov-Nambu-Jona-Lasinio (PNJL) models~\cite{Islam:2019tlo}, the color string percolation model~\cite{Sahu:2020mzo}, nonconformal holography~\cite{Grefa:2022sav}, the excluded-volume hadron resonance gas (EHRG) model~\cite{McLaughlin:2021dph}, and the dynamical quasiparticle model (DQPM)\cite{Saha:2017xjq,Chahal:2023khc,Pal:2023aai,Shaikh:2024ewz,Sharma:2024nkt}, as well as varying initial conditions~\cite{Liu:2018xae}, temperature~\cite{Achenbach:2023pba,Panday:2022sxb,Auvinen:2020mpc}, chemical potential~\cite{Sorensen:2021zme,He:2022kbc,He:2022yrk}, beam energy~\cite{Zabrodin:2020cfp}, and system rotation~\cite{Sahoo:2023xnu,Rath:2024vyq}, have been employed to study the properties of the QGP.
In recent years, data-driven approaches, particularly machine learning and Bayesian inference, have proven successful in extracting transport coefficients across system sizes~\cite{Nijs:2020roc,Bernhard:2019bmu,Moreland:2018gsh,JETSCAPE:2020mzn,JETSCAPE:2020shq,Heffernan:2023utr}, as summarized in Ref.~\cite{Achenbach:2023pba}. While these analyses depend on assumptions for many input parameters, a systematic study is crucial for the reliable extraction of real-time QGP properties.
Experimental and theoretical developments have also highlighted the importance of spin polarization (SP), which can be induced by strong vorticity generated in noncentral collisions. Since the first observation of hyperon SP at the RHIC~\cite{STAR:2017ckg,STAR:2019erd,STAR:2021beb,STAR:2022fan} and the LHC~\cite{ALICE:2019aid,ALICE:2021pzu}, SP has emerged as a promising probe of the nontrivial vortical structure of the QGP~\cite{Liang:2004ph,Florkowski:2017ruc,Wei:2018zfb,Florkowski:2019qdp,Fukushima:2020ucl,Weickgenannt:2020aaf, Shi:2020htn,Florkowski:2021wvk,Speranza:2021bxf,Das:2022azr,Becattini:2024uha,Palermo:2024tza,Becattini:2025oyi}, as reviewed in Refs.~\cite{Singh:2021yba,Chen:2024hki}, owing to its sensitivity to the transport properties of the medium~\cite{Palermo:2024tza,Serenone:2021zef}. Additionally, strong vorticity and magnetic fields in noncentral collisions significantly affect transport coefficients~\cite{Aung:2023pjf,Sahoo:2023xnu,Rath:2024vyq}. Therefore, elucidating the coupling between SP and transport and/or thermodynamic coefficients is vital for understanding the physical mechanisms of heavy-ion collisions, particularly concerning the quantum chromodynamics critical point.
In my previous work, I developed a kinetic theory framework showing that spin polarization can be induced in the distribution function via vorticity~\cite{Wei:2025rxh}. One found that the nonmonotonic behavior of $c_s^2$ and $\zeta/s$ may serve as a potential signature of the QCD critical point, though further investigation is necessary. In this study, I extend that framework by incorporating two contributions to SP in the distribution function: thermal vorticity and thermal shear. This allows us to understand the influence of SP and finite-volume effects on transport and thermodynamic coefficients. This work extract these coefficients from the modified distribution functions and analyze their behavior in O+O, Cu+Cu, Ag+Ag, and Au+Au collisions, aiming to identify potential signatures of the QCD critical point.

\section{Spin polarization in kinetic theory}
\label{method}

The spin-dependent phase-space distribution function in near-equilibrium state, can be parametrized as~\cite{Weickgenannt:2020aaf,Becattini:2013fla,Florkowski:2019gio}
\begin{eqnarray}\label{disf:101}
& & \tilde{f}_{mn}[r(\tau),p(\tau),s(\tau)]   \nonumber\\
&=& f[r(\tau),p(\tau)] \times \left[\delta_{mn}+\Pi[r(\tau),p(\tau)]\sigma_{mn}\right]\,.
\end{eqnarray}
Here, $m,n=1,2$ are spin indices, $f[r(\tau),p(\tau)]$ is the spinless phase-space distribution function, $s(\tau)$ denotes the spin degree of freedom in the phase-space distribution, $\sigma_{mn}$ is a three-vector composed of Pauli matrices, and $\Pi[r(\tau),p(\tau)]$ represents the spin-polarization three-vector.

The most common approach in polarization studies assumes that the spin polarization tensor is given by the thermal vorticity $\varpi_{\rho\sigma}$ (referred to as case I)~\cite{Becattini:2013fla}:
\begin{eqnarray}\label{ther:101}
\varpi_{\rho\sigma}=-\frac{1}{2}(\partial_{\rho}\beta_{\sigma}-\partial_{\sigma}\beta_{\rho})\,,
\end{eqnarray}
where $\beta_{\rho}=u_{\rho}/T$ with $u$ denoting the fluid four-velocity and $T$ the temperature.

The kinetic theory approach allows for corrections to the equilibrium distribution. In this case, such corrections are obtained from the kinetic equation in the relaxation-time approximation (RTA), which naturally leads to contributions from the thermal shear~\cite{Becattini:2021suc}
\begin{eqnarray}\label{ther:102}
\xi_{\rho\sigma}=\frac{1}{2}(\partial_{\rho}\beta_{\sigma}+\partial_{\sigma}\beta_{\rho})\,.
\end{eqnarray}
In particular, the thermal vorticity $\varpi_{\rho\sigma}$ and the thermal shear tensor $\xi_{\rho\sigma}$ correspond to the antisymmetric and symmetric parts of the fluid velocity gradient, respectively. The combined contribution of thermal vorticity and thermal shear (referred to as case II) is expressed as~\cite{Becattini:2021suc}
\begin{eqnarray}\label{ther:103}
\Omega_{\rho\sigma}=\varpi_{\rho\sigma}+2\tau\frac{p}{E}\xi_{\rho\sigma}\,,
\end{eqnarray}
where $p$ and $E$ denote the particle's momentum and energy, respectively, and $\tau$ is the unit normal vector.
The second term's factor is derived by coupling the antisymmetric stress-energy tensor with $\xi$, as demonstrated in ref.~\cite{Becattini:2021suc}.

For the study of spin polarization, the case I-type vector is expressed in terms of the thermal vorticity tensor: $\varpi^{\mu}= -1/2\epsilon^{\mu\rho\sigma\tau}\varpi_{\rho\sigma}u_{\tau}$.

The case II--type vector is expressed as the sum of thermal vorticity and thermal shear tensors: $\Omega^{\mu}=-1/2\epsilon^{\mu\rho\sigma\tau}\Omega_{\rho\sigma}u_{\tau}$.
Under the assumption of small thermal vorticity or thermal shear within the local thermodynamic equilibrium, a linear relationship arises between the average spin polarization vector and the thermal vorticity (or thermal vorticity plus thermal shear).
Specifically, for spin-$1/2$ particles, the polarization vector satisfies $\Pi^{\mu}\approx \varpi^{\mu}[2s(s+1)/3]$ (vorticity-induced polarization, VIP) or $\Pi^{\mu}\approx \Omega^{\mu}[2s(s+1)/3]$ (vorticity- and shear-induced polarization, VIP+SIP).

To highlight the contribution of spin polarization, one employs kinetic theory to compute the energy-momentum tensor. Taking into account the spin measure, the SP-dependent energy-momentum tensor is defined as
\begin{eqnarray}\label{tens:101}
\tilde{T}^{\mu\nu}(\tau,r) &=& \frac{1}{2}\int\frac{d^{3}p}{(2\pi)^{3}}\frac{p^{\mu}p^{\nu}}{p^{0}}\int dS\tilde{f}[r(\tau),p(\tau),s(\tau)] \nonumber\\
&=& \int\frac{d^{3}p}{(2\pi)^{3}}\frac{p^{\mu}p^{\nu}}{p^{0}}f[r(\tau),p(\tau)]  \nonumber\\
&\times& \frac{1}{2}\sum_{m,n=1}^{2}\int \left[\delta_{mn}+\Pi[r(\tau),p(\tau)]\sigma_{mn}\right]dS   \nonumber\\
&=& \int\frac{d^{3}p}{(2\pi)^{3}}\frac{p^{\mu}p^{\nu}}{p^{0}}f[r(\tau),p(\tau)]  \nonumber\\
&\times& \frac{1}{2}\sum_{m=1}^{2}\int \left[\delta_{mm}+\Pi[r(\tau),p(\tau)]\sigma_{mm}\right]dS \,.
\end{eqnarray}
Here, the invariant spin measure is parametrized as~\citep{Florkowski:2018fap}
\begin{eqnarray}\label{tens:102}
& & \frac{1}{2}\sum_{m=1}^{2}\int \left[\delta_{mm}+\Pi[r(\tau),p(\tau)]\sigma_{mm}\right]dS   \nonumber\\
&\approx& 2-\frac{\, \sinh[\chi\,\Pi[r(\tau),p(\tau)]]}{\chi\,\Pi[r(\tau),p(\tau)]}
\end{eqnarray}
with
\begin{eqnarray}\label{tens:103}
& & \int \langle\cdot\cdot\cdot\rangle dS=\frac{M}{\pi\chi}\int \langle\cdot\cdot\cdot\rangle d^{4}s\delta(s\cdot s+\chi^{2})\delta(p\cdot s) \,,
\end{eqnarray}
where $\chi^{2}=s(s+1)= {3}/{4}$, $M$ is the particle's mass, the factor $1/2$ denotes the normalization of the sum of particles and antiparticles (This work does not distinguish between particles and antiparticles, as the chemical potential is not introduced in this framework.) When spin is treated as an independent variable, disregarding contributions from thermal vorticity and thermal shear, the result of integrating the spin measure is $\int dS= 2$~\cite{Florkowski:2018fap,Bhadury:2020cop,Drogosz:2024gzv}. The distribution function sums over spin-up and spin-down particles. For comparison, the spinless energy-momentum tensor is given by
\begin{eqnarray}\label{tens:104}
T^{\mu\nu}(\tau,r) &=& \int\frac{d^{3}p}{(2\pi)^{3}}\frac{p^{\mu}p^{\nu}}{p^{0}}f[r(\tau),p(\tau)]\,.
\end{eqnarray}
The factor $d^{3}p/[(2\pi)^{3} p^{0}]$ is the Lorentz invariant integration measure in momentum space, where $p^{0}=\sqrt{m^{2}+p^{2}}$ is the on-shell energy of the particle.

By evaluating the energy-momentum tensor, transport and thermodynamic coefficients (TTCs) can be extracted, such as specific shear viscosity $\eta/s$, specific bulk viscosity $\zeta/s$, and mean free path $\lambda$, which are of central interest in QGP physics. The first-order, spinless TTCs for massive particles in kinetic theory are defined as (similar to ref.~\cite{Bozek:2009dw,Dusling:2011fd,AlvaradoGarcia:2023fsy})
\begin{eqnarray}\label{tran:101}
\eta/s &=& \frac{1}{15(\varepsilon+P)}\int\frac{d^{3}p}{(2\pi)^{3}}\frac{p^{4}}{p^{0}}f(1-f), \nonumber\\
\zeta/s &=& 15\eta/s\left(\frac{1}{3}-c_{s}^{2}\right)^{2}  \nonumber\\
&=& \frac{1}{(\varepsilon+P)}\left(\frac{1}{3}-c_{s}^{2}\right)^{2} \int\frac{d^{3}p}{(2\pi)^{3}}\frac{p^{4}}{p^{0}}f(1-f),  \nonumber\\
\lambda &\approx& \frac{5}{T}\eta/s     \nonumber\\
&=& \frac{1}{3(\varepsilon+P)T}\int\frac{d^{3}p}{(2\pi)^{3}}\frac{p^{4}}{p^{0}}f(1-f)\,.
\end{eqnarray}
Here, the local temperature is defined as $T=(\pi^{2}\varepsilon/72)^{1/4}$~\cite{Zhang:2015tta}, and $\varepsilon$ denotes the energy density. The energy density and pressure are obtained from \eq{tens:104}, specifically: $\varepsilon=T^{00},~P=\sqrt{(T^{11})^{2}+(T^{22})^{2}+(T^{33})^{2}}$. From the thermodynamic relation, we have the entropy density $s=(\varepsilon+P)/T$. The squared speed of sound is defined as the derivative of pressure with respect to the energy density at constant entropy: $c_{s}^{2}=(\partial P/\partial \varepsilon)_{s}$.

\begin{figure*}[tp]
\begin{center}
\includegraphics[width=0.720\textwidth]{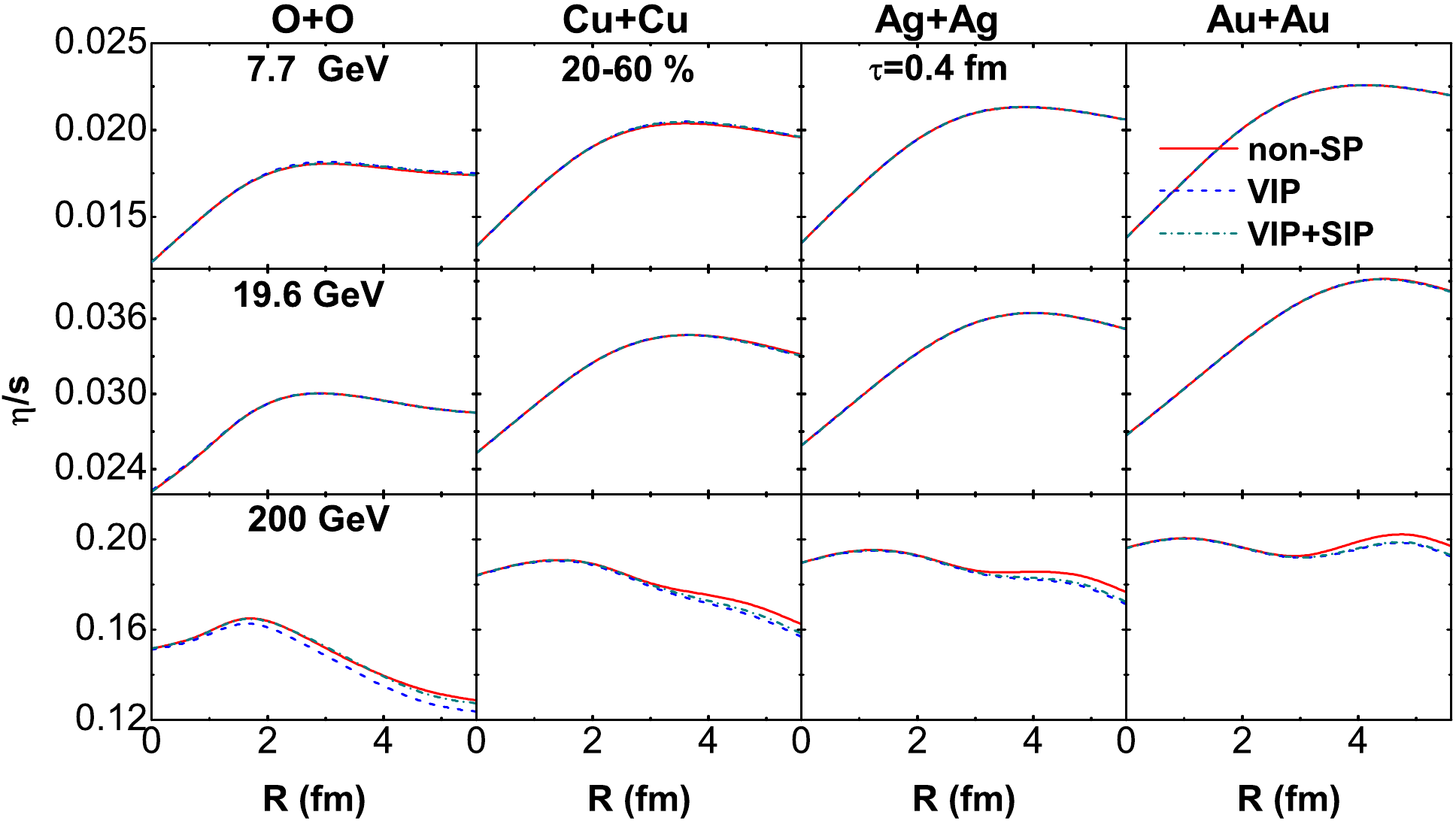}\\
\vspace{0.2cm}
\includegraphics[width=0.720\textwidth]{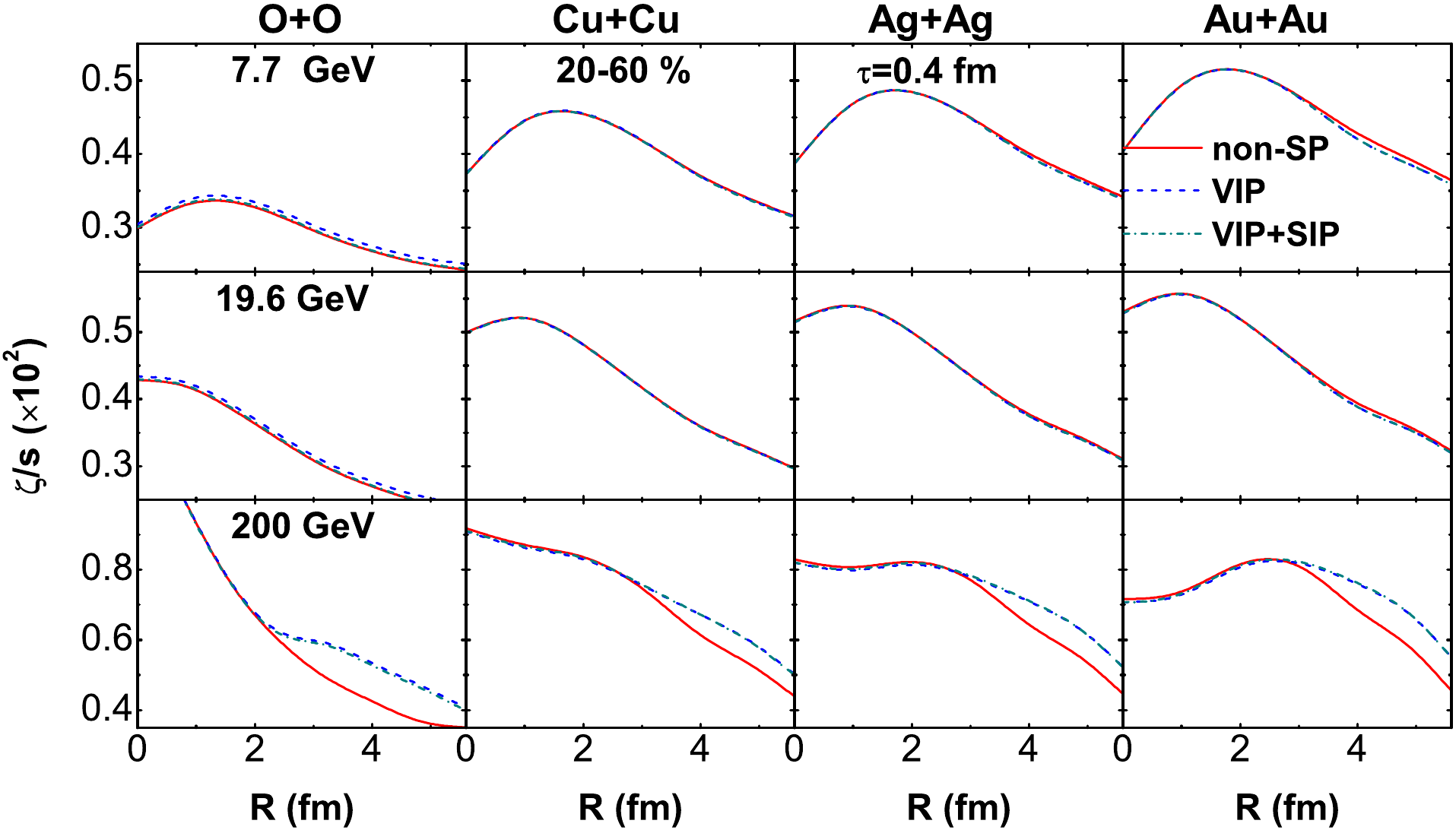}
\caption{(Color online)
Top panels: Comparison of the specific shear viscosity $\eta/s$ as a function of radius for different systems at $\sqrt{s_{NN}}$=7.7, 19.6, 200 GeV, using the VIP, VIP+SIP, and non-SP methods.
Bottom panels: Corresponding distributions for the specific bulk viscosity $\zeta/s$.
All results are shown at proper time $\tau=0.4$ fm for each system.
}
\label{fig1}
\end{center}
\end{figure*}

\begin{figure*}[tp]
\begin{center}
\includegraphics[width=0.720\textwidth]{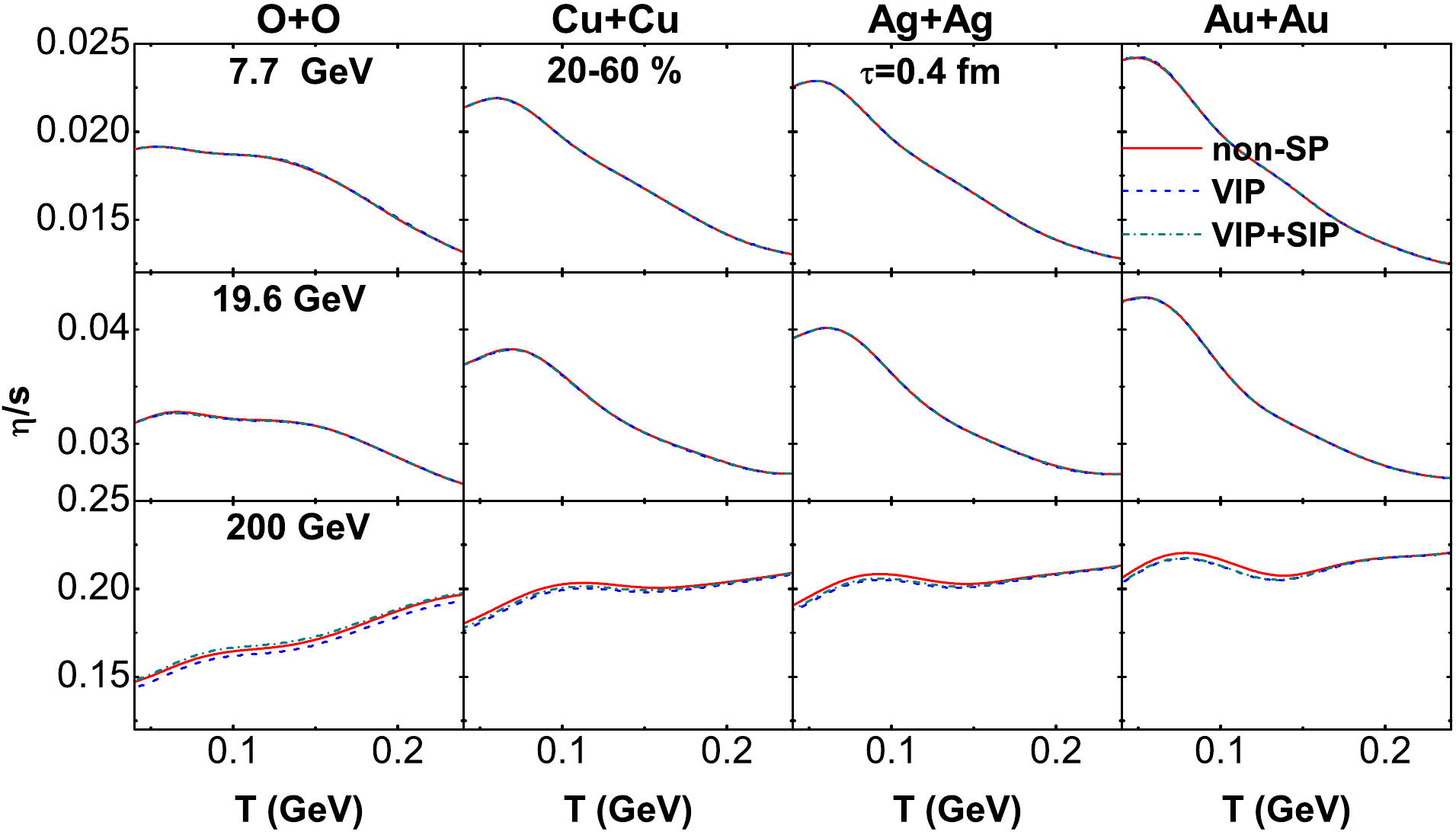}\\
\vspace{0.2cm}
\includegraphics[width=0.720\textwidth]{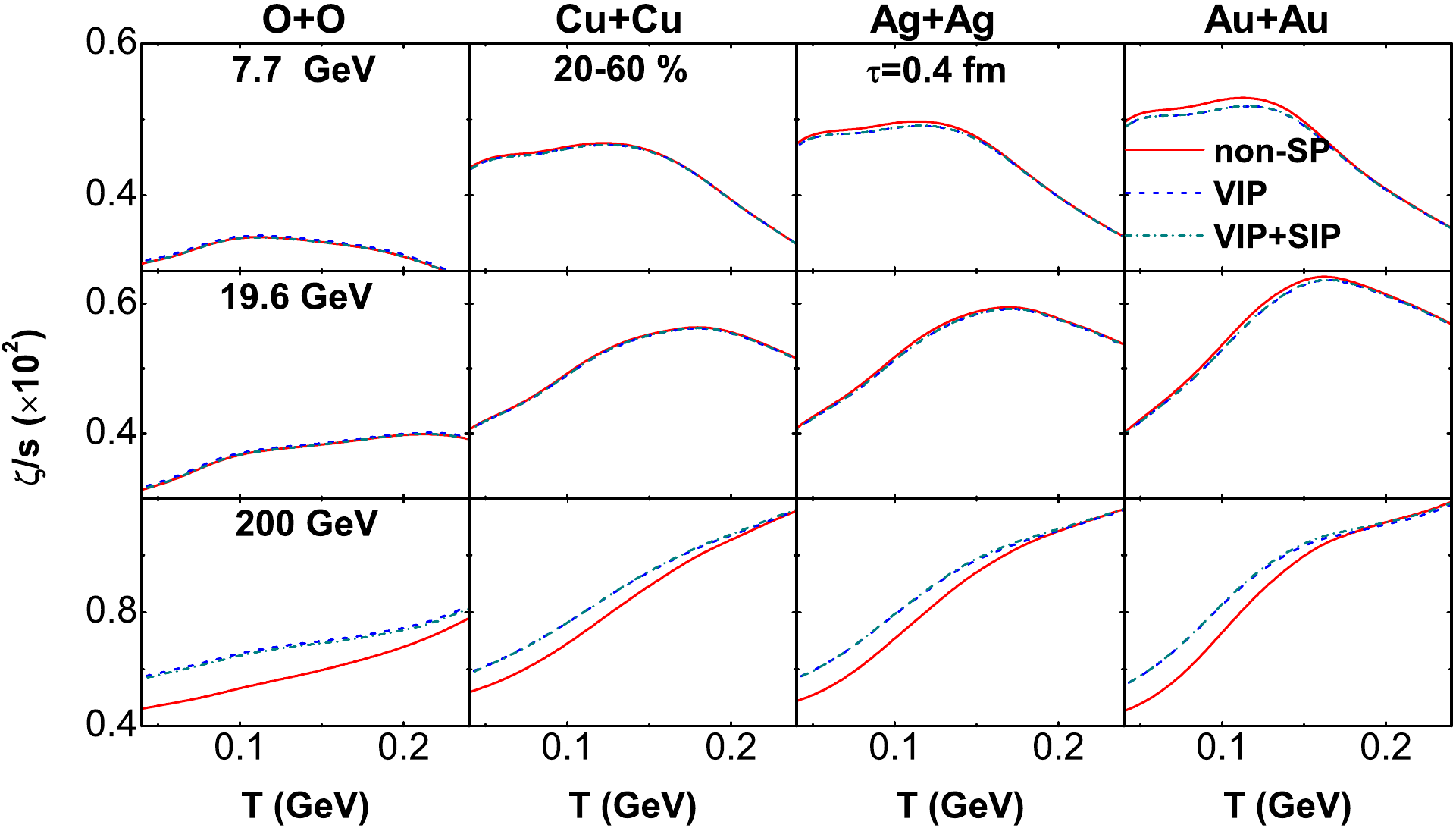}
\caption{(Color online)
Top panels: Comparison of the specific shear viscosity $\eta/s$ as a function of temperature for different systems at $\sqrt{s_{NN}}$=7.7, 19.6, and 200 GeV, using the VIP, VIP+SIP, and non-SP methods.
Bottom panels: Corresponding distributions of the specific bulk viscosity $\zeta/s$.
All results are shown at proper time $\tau=0.4$ fm for each system.
}
\label{fig2}
\end{center}
\end{figure*}

Similarly, the first-order spin-polarization-dependent TTCs for massive particles are defined as:
\begin{eqnarray}\label{tran:102}
\tilde{\eta}/\tilde{s} &=& \frac{1}{15(\tilde{\varepsilon}+\tilde{P})}\int\frac{d^{3}p}{(2\pi)^{3}}\frac{p^{4}}{p^{0}}dS\tilde{f}(1-\tilde{f}), \nonumber\\
\tilde{\zeta}/\tilde{s} &=& \frac{1}{(\tilde{\varepsilon}+\tilde{P})}\left(\frac{1}{3}-\tilde{c}_{s}^{2}\right)^{2} \int\frac{d^{3}p}{(2\pi)^{3}}\frac{p^{4}}{p^{0}}dS\tilde{f}(1-\tilde{f}),  \nonumber\\
\tilde{\lambda} &\approx& \frac{1}{3(\tilde{\varepsilon}+\tilde{P})T}\int\frac{d^{3}p}{(2\pi)^{3}}\frac{p^{4}}{p^{0}}dS\tilde{f}(1-\tilde{f})\,,
\end{eqnarray}
where $\tilde{\varepsilon}=\tilde{T}^{00},~\tilde{P}=\sqrt{(\tilde{T}^{11})^{2}+(\tilde{T}^{22})^{2}+(\tilde{T}^{33})^{2}}$, and $\tilde{c}_{s}^{2}=(\partial \tilde{P}/\partial \tilde{\varepsilon})_{s}$.
The vorticity tensor, the shear tensor, and the viscosity coefficients are dissipative quantities related to the relaxation time. To quantitatively compare the contributions of the two sources of spin polarization, I compute the specific viscosities as illustrated in Figs.~\ref{fig1} and \ref{fig2}. In \Fig{fig1}, one presents the radial dependence of the specific shear viscosity $\eta/s$ (top panels) and the specific bulk viscosity $\zeta/s$ (bottom panels) for various systems at $\sqrt{s_{NN}} = 7.7$, 19.6, and 200~GeV. The results are shown at an early proper time, $\tau = 0.4$~fm.
Comparing three different scenarios--VIP, VIP+SIP, and non-SP--one observes that the specific viscosities are significantly influenced by the inclusion of spin polarization. The splitting effect in both $\eta/s$ and $\zeta/s$ becomes more pronounced with higher collision energies, while showing a weak dependence on the size of the system. Across the radial range, spin polarization tends to suppress $\eta/s$ monotonically with increasing energy. In contrast, the impact on $\zeta/s$ exhibits a sign reversal: it is suppressed at lower energies except for O+O systems, but enhanced at higher energies.
This opposite behavior arises from the different physical roles of the two viscosities. Shear viscosity reflects momentum transport and is mainly influenced by Coriolis forces in rotating systems~\citep{Aung:2023pjf}, while bulk viscosity captures pressure variations associated with local expansion or compression. Consequently, spin polarization contributes differently to these transport coefficients.
In \Fig{fig2}, one shows analogous results but as functions of temperature. Across the temperature range, spin polarization suppresses $\eta/s$. The behavior of $\zeta/s$ remains consistent with that of \Fig{fig1}, showing suppression at lower energies and enhancement at higher energies.
It should be noted that in the lower-energy O-O collisions with $\tau=0.4$ fm, the system may not have yet reached thermal equilibrium, and the applicability of spin polarization in this regime requires careful consideration.

From the results across different system sizes shown in Figs.~\ref{fig1} and \ref{fig2}, it is evident that the contributions of VIP and SIP to the distribution function are energy dependent. Specifically, VIP and VIP+SIP dominate the polarization coefficient across most energy ranges, while SIP alone has a negligible impact at lower energies but becomes relatively significant at higher energies. These differences stem primarily from the underlying mechanisms: spin polarization induced by thermal vorticity (VIP) is independent of particle velocity, whereas that induced by thermal shear (SIP) is closely tied to particle motion. As a result, SIP contributions are proportional to the system's topological quantum number, leading to distinct macroscopic effects.
In this work, TTCs are studied in O+O, Cu+Cu, Ag+Ag, and Au+Au collisions at 20-60 $\%$ centrality using a multi-phase transport (AMPT) model~\cite{Lin:2004en}. No multiplicity filter is applied; instead, the centrality is fixed, yielding a broad range of multiplicities from low to high. The default collision probability parameters in AMPT are used: $a=0.5$, $b=0.9$~GeV$^{-2}$, $\alpha_s = 0.33$, and $\mu = 3.2$~\citep{MacKay:2022uxo,Zhang:2023lzv}.
As noted in the previous work, the correlation length diverges near the QCD critical point~\cite{Stephanov:1998dy}, leading to singular behavior in several thermodynamic quantities that may influence the QGP signatures. The probability of observing such critical effects increases if the freeze-out trajectory in the $\mu$-$T$ plane lies sufficiently close to the critical point.

To investigate the impact of spin polarization on transport and thermodynamic properties, one extracts the TTC within the kinetic theory framework, as defined in Eqs.~(\ref{tran:101}) and (\ref{tran:102}).
Spin polarization is considered only during the partonic stage and is assumed to arise solely from thermal vorticity and/or thermal shear, neglecting other possible sources such as magnetic fields.

This work adopt natural units with $k_B=c=\hbar=1$.

\section{Numerical results and discussions}
\label{results}

\begin{figure*}[tp]
\begin{center}
\includegraphics[width=0.720\textwidth]{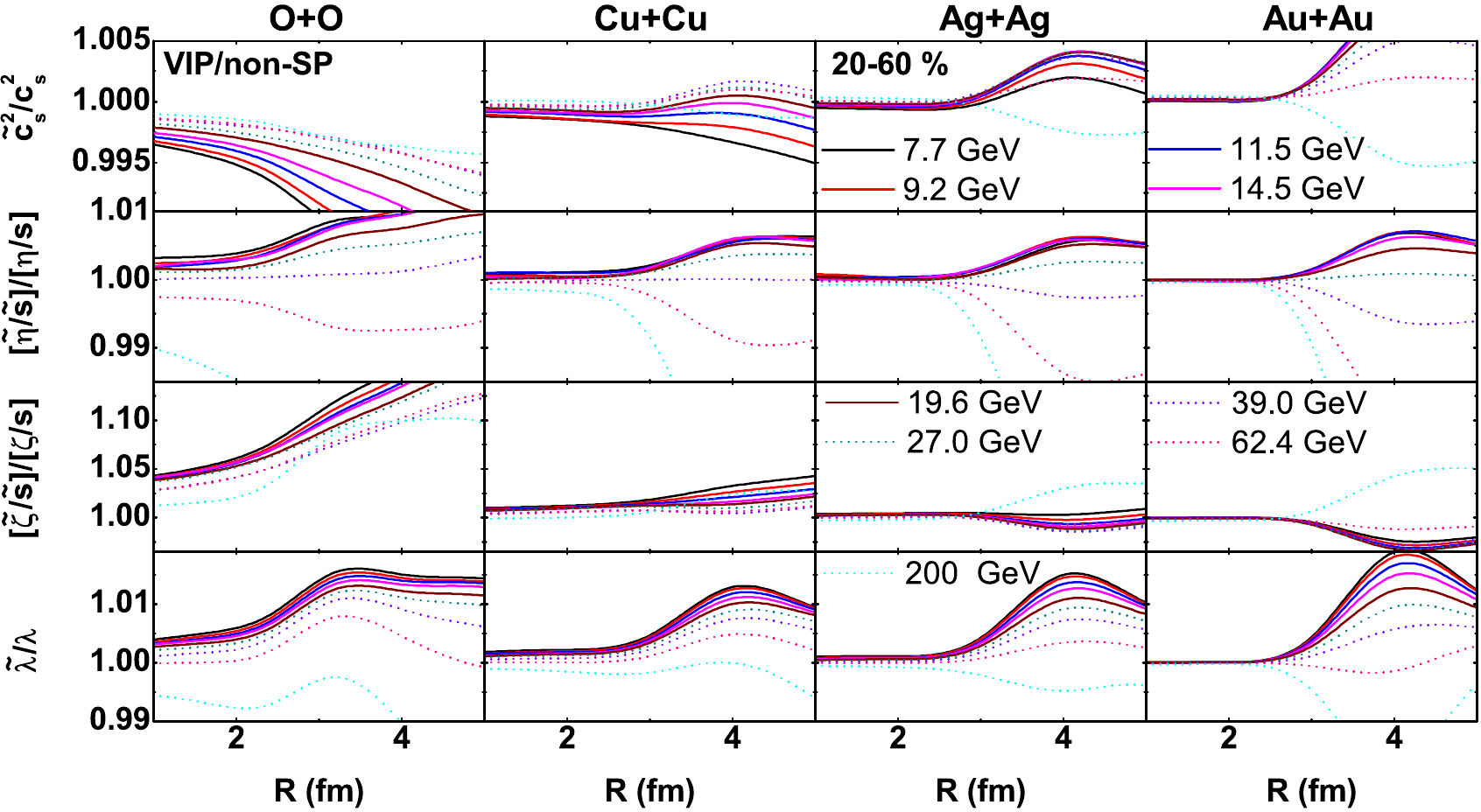}\\
\vspace{0.2cm}
\includegraphics[width=0.720\textwidth]{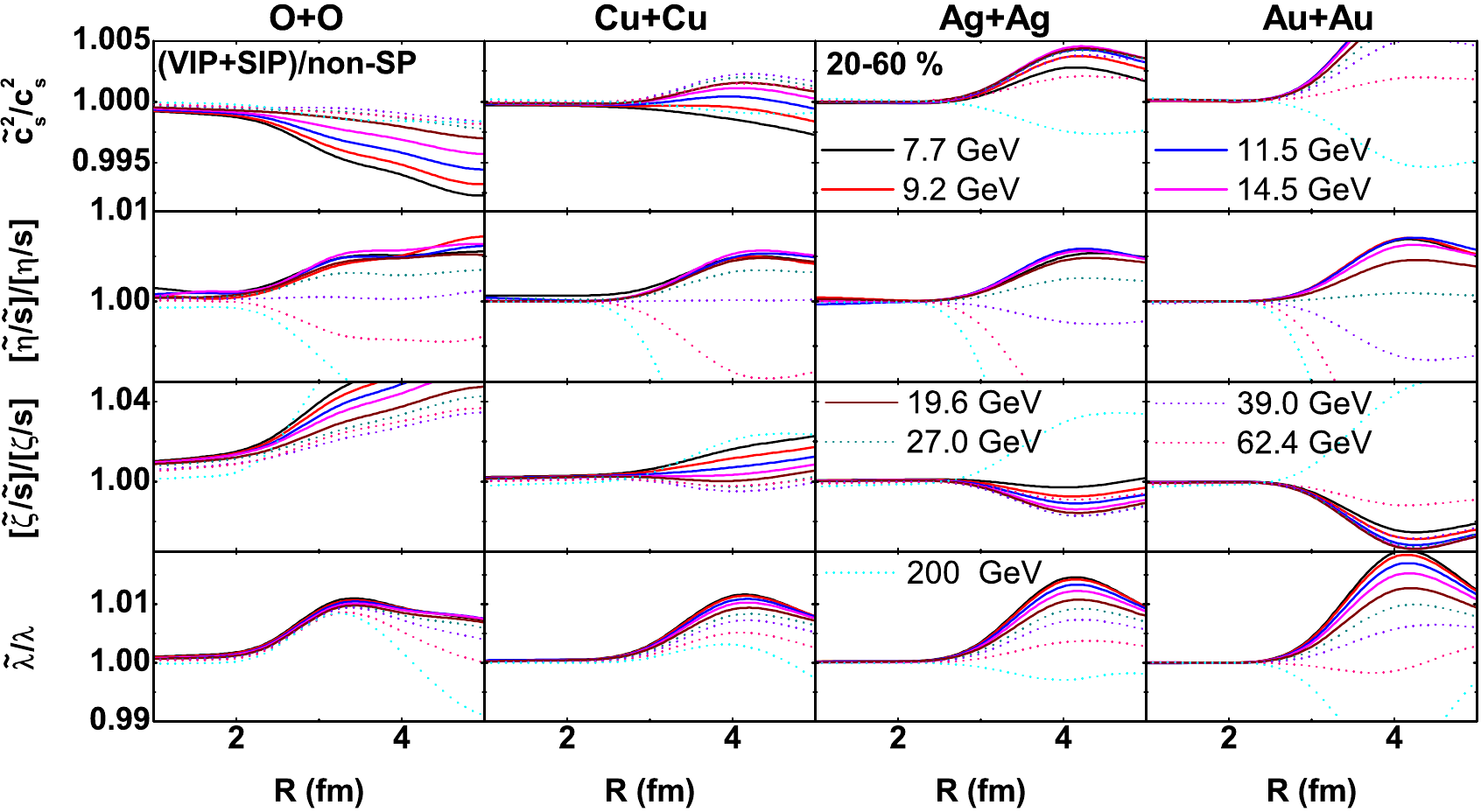}
\caption{(Color online)
Top panels: Ratios of the coefficients ($c_{s}^{2}$, $\eta/s$, $\xi/s$, $\lambda$) between the VIP and non-SP methods, shown as functions of radius for various systems at $\sqrt{s_{NN}}$ = 7.7-200~GeV.
Bottom panels: Ratios of the same coefficients between the VIP+SIP and non-SP methods, also as functions of radius for the same systems and energy range.
All results are obtained by summing over the proper time interval from $\tau = 0$ to 6 fm. 
See Supplemental Material~\cite{Wei:2025sup} for a more complete data presentation.
}
\label{fig3}
\end{center}
\end{figure*}

\begin{figure*}[tp]
\begin{center}
\includegraphics[width=0.720\textwidth]{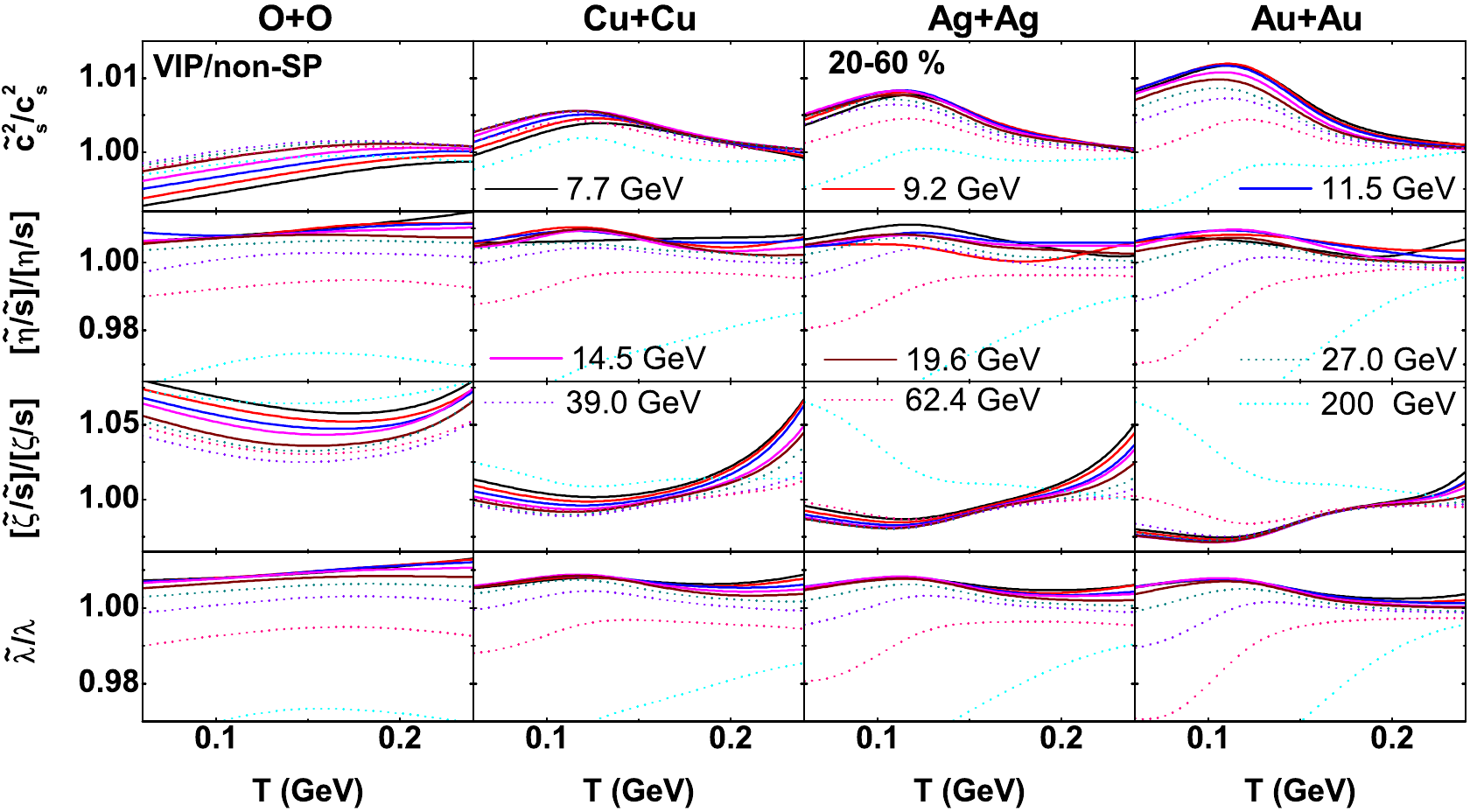}\\
\vspace{0.2cm}
\includegraphics[width=0.720\textwidth]{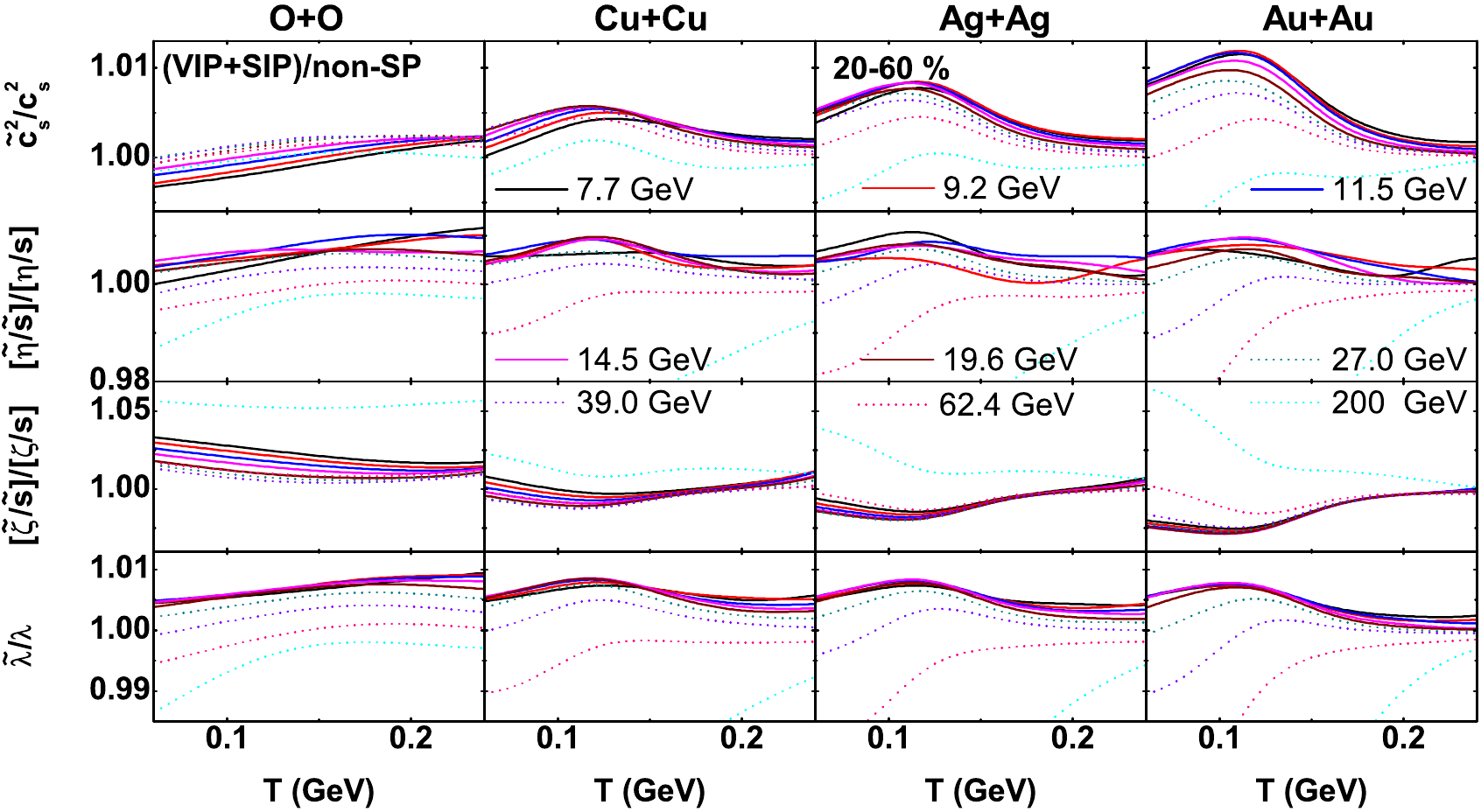}
\caption{(Color online)
Top panels: Comparison of the coefficients ($c_{s}^{2}$, $\eta/s$, $\xi/s$, $\lambda$) between the VIP and non-SP methods, shown as functions of temperature for various systems at $\sqrt{s_{NN}}$ = 7.7-200~GeV.
Bottom panels: Same comparison as above, but between the VIP+SIP and non-SP methods.
All results are obtained by summing over the proper time interval from $\tau = 0$ to 6 fm.
See Supplemental Material~\cite{Wei:2025sup} for a more complete data presentation.
}
\label{fig4}
\end{center}
\end{figure*}

Having introduced the kinetic theory framework, one now presents numerical results for the transport and thermodynamic coefficients (TTCs). A more detailed numerical analysis will be reported in a forthcoming study.

To examine the influence of the two sources of spin polarization on phase-space distributions, one show the volume-dependent ratios of the TTCs in \Fig{fig3} and the temperature-dependent ratios in \Fig{fig4}.

In the top panels of \Fig{fig3}, one shows the radial dependence of $c_{s}^{2}$, $\eta/s$, $\zeta/s$, and $\lambda$, comparing the VIP and non-SP methods for various systems at $\sqrt{s_{NN}} = 7$-200~GeV. The bottom panels show analogous distributions for the ratio between VIP+SIP and non-SP methods. In both cases, the results show a nonmonotonic energy dependence of the ratios $\tilde{c}_{s}^{2}/c_{s}^{2}$ and $[\tilde{\zeta}/\tilde{s}]/[\zeta/s]$ at $R \geq 3$~ fm (except for O+O systems in VIP/non-SP case), whereas the ratios $[\tilde{\eta}/\tilde{s}]/[\eta/s]$ and $\tilde{\lambda}/\lambda$ show a monotonic energy dependence.

As noted in the previous work~\cite{Wei:2025rxh}, spin polarization increases with radial distance due to the rising orbital angular momentum, leading to a shift of the critical signal away from the collision center.

Figure~\ref{fig4} presents distributions similar to those in \Fig{fig3}, but as functions of temperature. The results are integrated over the proper time interval $\tau = 0$-6~fm. In both VIP/non-SP and (VIP+SIP)/non-SP cases, a nonmonotonic energy dependence is observed for $\tilde{c}_{s}^{2}/c_{s}^{2}$ and $[\tilde{\zeta}/\tilde{s}]/[\zeta/s]$ at $T \leq 0.17$~GeV, while the energy dependence of $[\tilde{\eta}/\tilde{s}]/[\eta/s]$ and $\tilde{\lambda}/\lambda$ remains monotonic.
In both cases, both $[\tilde{\eta}/\tilde{s}]/[\eta/s]$ and $\tilde{\lambda}/\lambda$ exhibit fluctuations at lower energies. These fluctuations demonstrate a very weak dependence on the system size.

Synthesizing the findings of Figs.~\ref{fig3} and \ref{fig4}, the observed nonmonotonic behavior may signal the presence of a first-order phase transition or a nearby critical point, likely associated with the transition from the hadronic phase to the QGP phase. This transition effectively reduces the degree-of-freedom available for spin polarization. Moreover, both figures indicate that the monotonic behavior of $[\tilde{\eta}/\tilde{s}]/[\eta/s]$ and $\tilde{\lambda}/\lambda$ is weakly dependent on system size. Interestingly, the nonmonotonic behavior in $\tilde{c}_{s}^{2}/c_{s}^{2}$ and $[\tilde{\zeta}/\tilde{s}]/[\zeta/s]$ also shows only a weak dependence on system size.

It is important to note that spin polarization characterizes a nondissipative property in thermodynamic equilibrium, while specific viscosities are dissipative quantities relevant to the nonequilibrium evolution of the system. Therefore, the contribution of spin polarization to the viscosity coefficients may reflect phase-transition-associated scale invariance.

\begin{figure*}[tp]
\begin{center}
\includegraphics[width=0.720\textwidth]{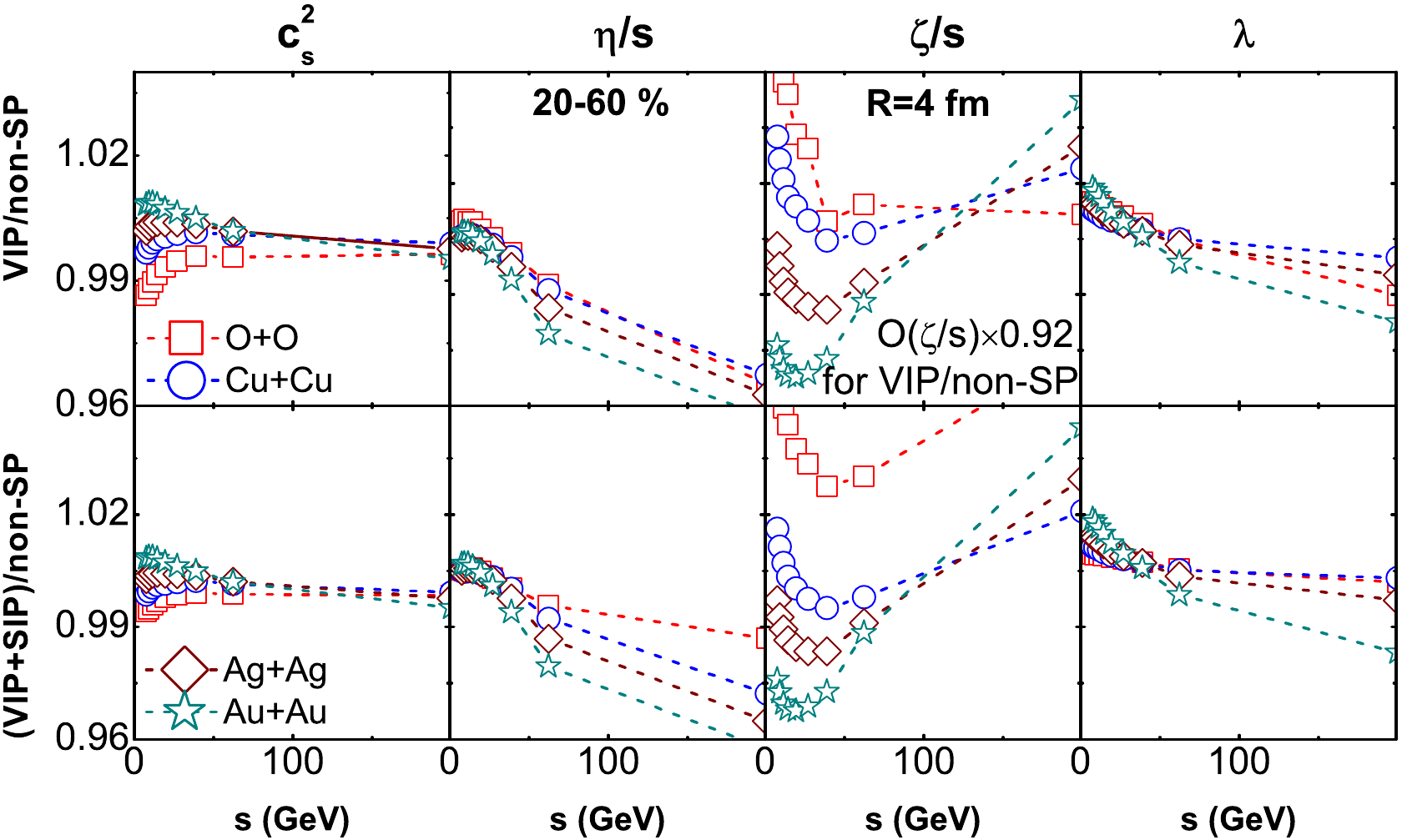}\\
\vspace{0.2cm}
\includegraphics[width=0.720\textwidth]{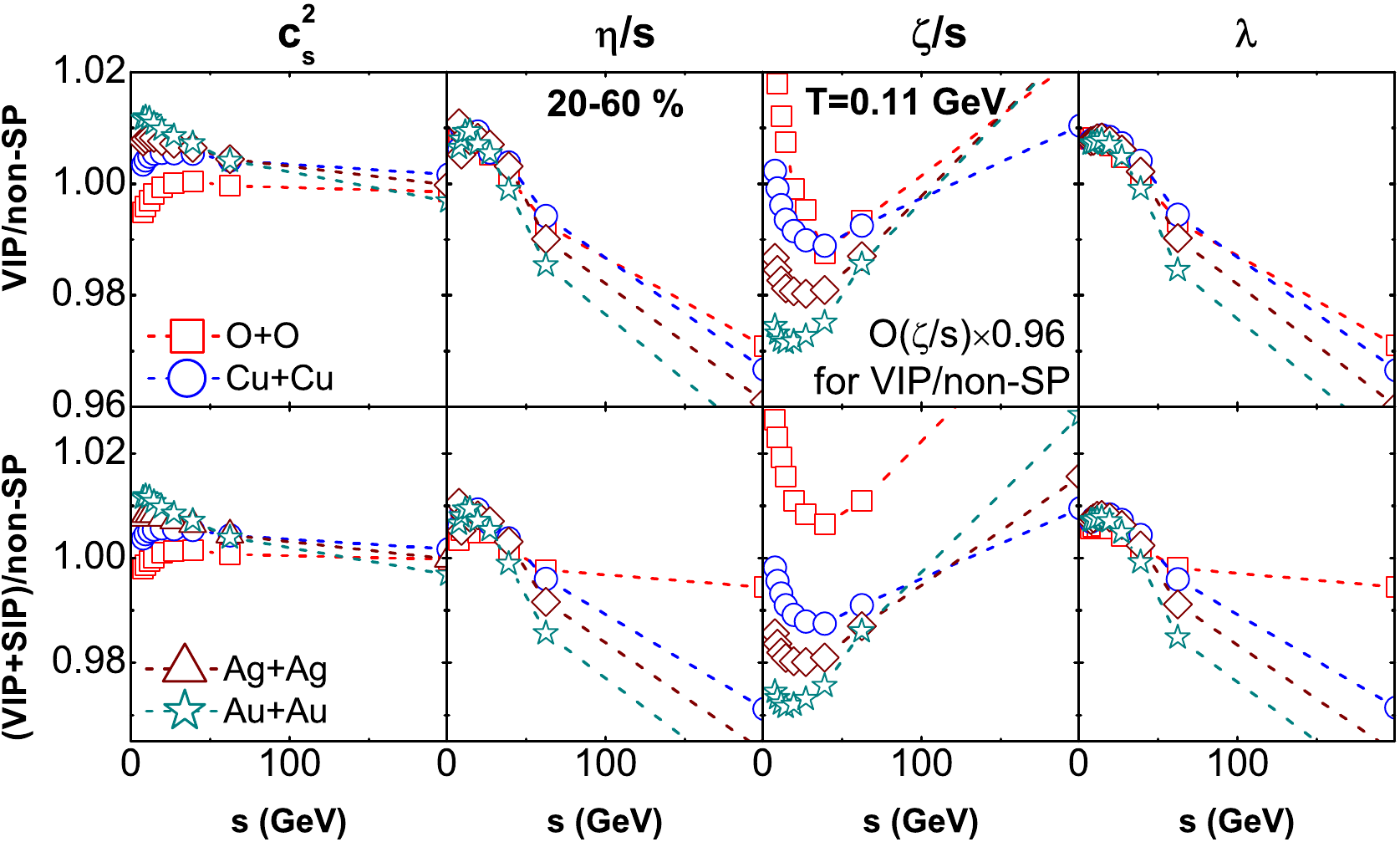}
\caption{(Color online)
Top panels: Ratios of the coefficients ($c_{s}^{2}$, $\eta/s$, $\xi/s$, $\lambda$) between VIP and non-SP methods, and between VIP+SIP and non-SP methods, shown as functions of collision energy for different systems at $R = 4$~fm. These results are extracted from \Fig{fig3}.
Bottom panels: Same comparisons as above, but shown as functions of collision energy at fixed temperature $T = 0.110$~GeV. These results are extracted from \Fig{fig4}.
}
\label{fig5}
\end{center}
\end{figure*}

In \Fig{fig5}, one presents the $R$-$\sqrt{s_{NN}}$ and $T$-$\sqrt{s_{NN}}$ distributions of the ratios $\tilde{c}_{s}^{2}/c_{s}^{2}$, $[\tilde{\eta}/\tilde{s}]/[\eta/s]$, $[\tilde{\zeta}/\tilde{s}]/[\zeta/s]$, and $\tilde{\lambda}/\lambda$. The top panels of \Fig{fig5} show the $R$-$\sqrt{s_{NN}}$ distributions with coefficients extracted from \Fig{fig3}, while the bottom panels display the $T$-$\sqrt{s_{NN}}$ distributions, based on the coefficients of \Fig{fig4}.

By comparing results between VIP and non-SP methods, as well as between VIP+SIP and non-SP methods, one observes that the energy dependence of $[\tilde{\eta}/\tilde{s}]/[\eta/s]$ and $\tilde{\lambda}/\lambda$ is monotonic in both radius and temperature distributions. This behavior exhibits a weak dependence on the system size. In contrast, the nonmonotonic behavior of $\tilde{c}_{s}^{2}/c_{s}^{2}$ and $[\tilde{\zeta}/\tilde{s}]/[\zeta/s]$  shows a significant dependence on system size.

These observations imply that the nonmonotonic behaviors follow relatively robust scaling laws across different systems. The emergence of such scaling laws is a characteristic feature of critical phenomena, suggesting that they may serve as potential signatures of the QCD critical point. As illustrated in \Fig{fig5}, the coefficient ratios in the O + O system are consistently higher than those in the Au+Au system. This difference can be primarily attributed to the stronger spin polarization generated in smaller systems, as previously reported in Ref.~\cite{Alzhrani:2022dpi}.

The observation of nonmonotonic behavior aligns with predictions from ideal hydrodynamics + UrQMD simulations on the information entropy of net proton distributions~\cite{Deng:2024abo}, as well as with RHIC measurements of high-order cumulants, scaling exponents, and light nuclei yield ratios~\cite{STAR:2020tga,STAR:2021fge,STAR:2022etb,STAR:2022hbp}. These results suggest that the critical point of QCD is likely limited to the energy range $\sqrt{s_{NN}} = 19.6$-27~GeV.

Furthermore, the VIP and VIP+SIP methods both indicate that vorticity and shear effects are more prominent near the QCD critical point than in other regions, consistent with predictions from the NJL model~\cite{Zhang:2020hha}. This supports the possible existence of a first-order phase transition and a nearby critical point around $\sqrt{s_{NN}} \sim 27$~GeV. Given that momentum-dependent SIP competes with momentum-independent VIP, the relaxation dynamics near the critical point becomes more intricate. Consequently, the critical behavior spans a broader energy range when SIP contributions are considered.

Interestingly, lattice QCD calculations suggest that rotation shifts the critical point of QCD toward a higher baryon chemical potential~\cite{Braguta:2025ddq}, a result that contrasts with predictions from the Einstein-Maxwell-Dilaton (EMD) model~\cite{Chen:2020ath}. However, in this study, neither rotational nor shear effects lead to a shift in the peak or valley positions of the observables. In other words, these effects do not appear to alter the location of the QCD critical point. This discrepancy likely arises from model assumptions: both the lattice and EMD studies assume rigid rotation, whereas this work incorporates nonrigid rotational and shear effects inherent in relativistic heavy-ion collisions.

\begin{figure}[t]
\begin{center}
\includegraphics[width=0.480\textwidth]{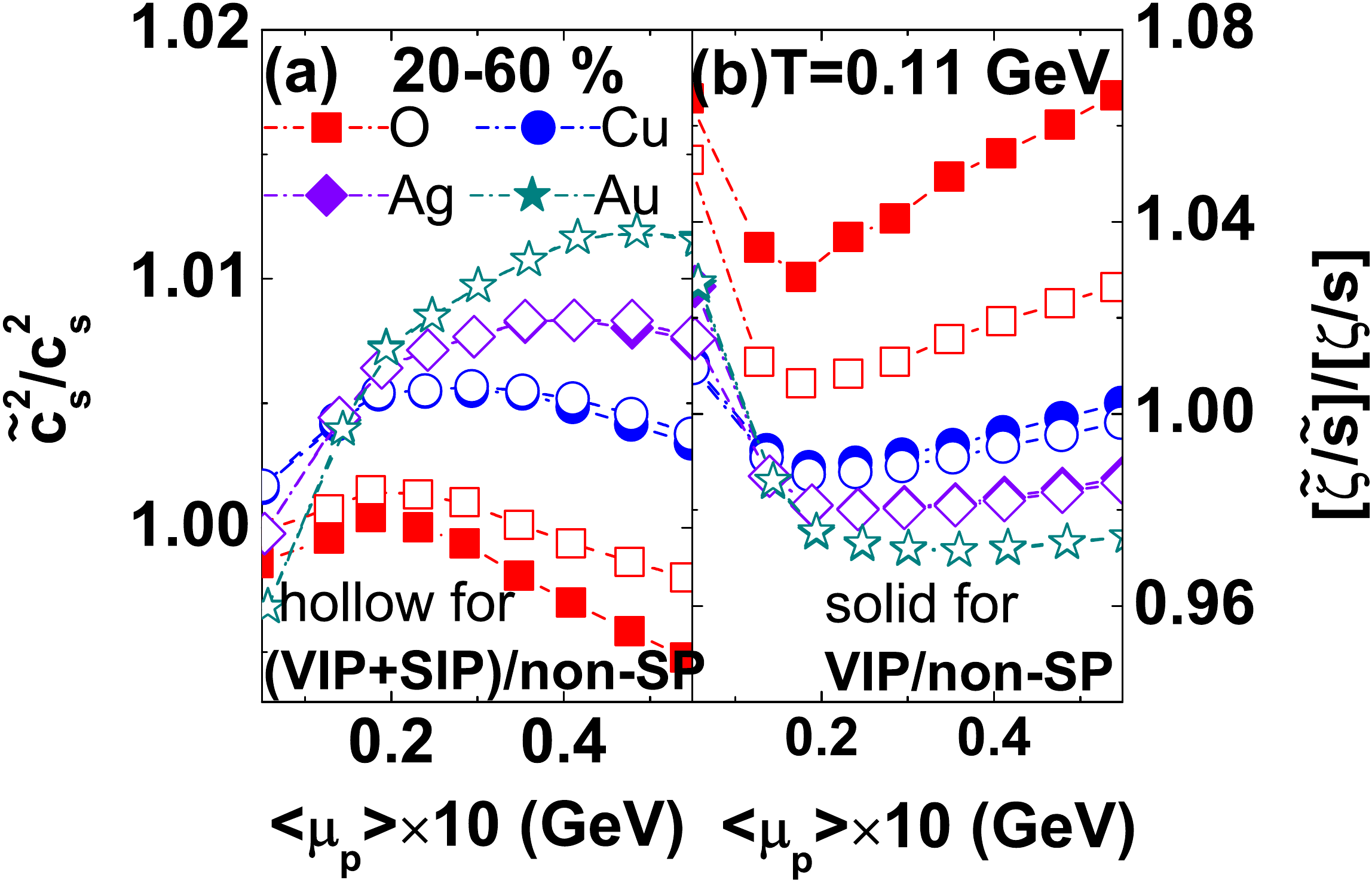}
\caption{(Color online)
Left panel: Ratios of $\tilde{c}_{s}^{2}/c_{s}^{2}$ between VIP and non-SP methods (solid symbols), and between VIP+SIP and non-SP methods (hollow symbols), shown as functions of the parton chemical potential $\langle \mu_{p} \rangle$.
Right panel: Ratios of $[\tilde{\zeta}/\tilde{s}]/[\zeta/s]$ for the same comparisons, also plotted as functions of $\langle \mu_{p} \rangle$.
}
\label{fig6}
\end{center}
\end{figure}

To investigate QCD properties at finite chemical potential, I also introduce the parton chemical potential, defined as
\begin{eqnarray}\label{chem:101}
\langle\mu_{p}\rangle &=& \frac{\int\mu_{p}(\tau)d\tau}{\int d\tau}\,,
\end{eqnarray}
where $\mu_{p}(\tau)=T_{c}/2*log[N_{p}(\tau)/N_{\bar{p}}(\tau)]$, following the approach in Ref.~\cite{Yu:2014epa}. Here, the total number of partons is defined as $N_{p}(\tau)=N_{u}(\tau)+N_{d}(\tau)+N_{s}(\tau)$, and the total number of anti-partons as $N_{\bar{p}}(\tau)=N_{\bar{u}}(\tau)+N_{\bar{d}}(\tau)+N_{\bar{s}}(\tau)$, where the subscripts denote the quark flavors. The critical temperature is taken as $T_{c}$=0.110 GeV. Following my previous work~\cite{Wei:2025rxh}, one extracts the parton chemical potential, which exhibits monotonic dependence on the collision energy.

In \Fig{fig6}, the ratios $\tilde{c}_{s}^{2}/c_{s}^{2}$ and $[\tilde{\zeta}/\tilde{s}]/[\zeta/s]$ are extracted from the top panels of \Fig{fig5}. For both VIP/non-SP and (VIP+SIP)/non-SP cases, the nonmonotonic behavior of these ratios restricts the inflection point to $\langle \mu_{p} \rangle \sim$ [0.018, 0.029]~GeV, corresponding to collision energies $\sqrt{s_{NN}} \sim$ [19.6, 39.0]~GeV.

The spin polarization induced by thermal shear only suppresses or enhances the magnitude of the coefficient ratios, without shifting the locations of their peaks or valleys. This observation suggests that the nonmonotonic behavior is primarily driven by spin polarization arising from thermal vorticity.

\section{Summary}
\label{summary}

This work employ the AMPT model to simulate the space-time evolution of noncentral light- and heavy-ion collisions. Two sources of spin polarization are considered in the calculations: thermal vorticity, induced by angular momentum, and thermal shear, arising from the velocity gradient.

Using the kinetic theory framework, one finds that the transport and thermodynamic coefficients, including the speed of sound squared $c_{s}^{2}$, specific shear viscosity $\eta/s$, specific bulk viscosity $\zeta/s$, and the mean free path $\lambda$, are significantly affected by both VIP and VIP+SIP contributions.

In the simulation, the impact of spin polarization on $\eta/s$ and $\lambda$ exhibits a monotonic dependence on collision energy. In contrast, its effect on $c_{s}^{2}$ and $\zeta/s$ shows a nonmonotonic dependence, particularly at a fixed radius $R = 4$~fm or at the critical temperature $T = 0.110$~GeV. Spin polarization induced by thermal shear only suppresses or enhances the magnitude of the coefficient ratios, without shifting the positions of their peaks or valleys.

Moreover, the contribution of spin polarization exhibits scale-dependent behavior across different system sizes, which may serve as an indicator of the QCD critical point.

Note that this study considers only the effects of VIP and/or VIP+SIP, while neglecting other possible sources of spin polarization, such as magnetic fields. Additionally, this work is based on kinetic theory and does not include the self-consistent evolution of the underlying microscopic dynamics. A more comprehensive model should incorporate these effects.

\begin{acknowledgments}
This work is supported by NSFC through Grant No.~12105057, and Guangxi NSF (China) through Grant Nos.~2023GXNSFAA026020, 2019GXNSFBA245080.
\end{acknowledgments}

\bibliographystyle{apsrev4-1}
\bibliography{References}

\end{document}